\documentclass{iau} 
\usepackage{graphicx}
\usepackage{subcaption}
\usepackage{hyperref}
\usepackage{sidecap}
\sidecaptionvpos{figure}{t}
\usepackage{wrapfig}
\usepackage{color}

\pubyear{2017}
\volume{331}  %% insert here IAU Symposium No.
\jname{SN 1987A, 30 years later – Cosmic Rays and Nuclei from Supernovae and their aftermaths}
\editors{A. Marcowith, G. Dubner, A. Ray, A. Bykov,\& M. Renaud, eds.}
\title[Investigation of 3C 397] %% give here short title %%
{Investigating the region of \\ 3C 397 in High Energy Gamma rays}

\author[P. Bhattacharjee, P. Majumdar, T. Ergin, L. Saha \& P. S. Joarder]   %% give here short author list %%
{Pooja Bhattacharjee$^1$, Pratik Majumdar$^2$, Tulun Ergin$^3$, Lab Saha$^4$ 
 \and Partha S. Joarder$^5$}
\affiliation{$^1$Centre for Astroparticle Physics and Space Science, Bose Institute,
               Block EN, Sector V, Salt Lake, Kolkata 700091, India, \\ Department of Physics, Bose Institute, 93/1 A.P.C. Road, Kolkata 700009, India \\ email: {\tt pooja.bhattacharjee@jcbose.ac.in} \\[\affilskip]
$^2$Saha Institute of Nuclear Physics, HBNI, 1/AF Bidhannagar, Kolkata 700064 \\email: {\tt pratik.majumdar@saha.ac.in} \\[\affilskip]
$^3$TUBITAK Space Technologies Research Institute, ODTU Campus, 06800 Ankara, Turkey \\ email: {\tt ergin.tulun@gmail.com} \\[\affilskip]
$^4$Universidad de Complutense, E-28040 Madrid, Spain \\ email: {\tt labsaha@ucm.es} \\[\affilskip]
$^5$Centre for Astroparticle Physics and Space Science, Bose Institute, Block EN, Sector V, Salt Lake, Kolkata 700091, India, \\ Department of Physics, Bose Institute, 93/1 A.P.C. Road, Kolkata 700009, India \\ email: {\tt partha@jcbose.ac.in}}

\begin{document}
\maketitle
%%%ABSTRACT
\begin{abstract}
We investigate the supernova remnant (SNR) 3C 397 and its neighboring pulsar PSR J1906+0722 in high energy gamma rays by using nearly six years of archival data of {\it Large Area Telescope} on board {\it Fermi Gamma Ray Space Telescope} (Fermi-LAT). The off-pulse analysis of gamma-ray flux from the location of PSR J1906+0722 reveals an excess emission which is found to be very close to the radio location of 3C 397. Here, we present the preliminary results of this gamma-ray analysis of 3C 397 and PSR J1906+0722.
%%%KEYWORDS
\keywords{Supernova Remnant, 3C 397, Pulsar, PSR J1906+0722, High Energy Gamma Rays}
\end{abstract}

\vspace{-0.5cm}
%%%MAIN TEXT
%%Introduction
\section{Introduction}
3C 397 (G41.1-0.3) is a radio and X-ray bright young Galactic Supernova Remnant (SNR) located 1$^{\circ}$ away from the location of PSR J1906+0722 (\cite{Safi-Harb2000}). The estimated age of 3C 397 is $\sim$ 5300 year (\cite{Safi-Harb2005}). Whether 3C 397 is a ``mixed-morphology" SNR (\cite{Safi-Harb2005}) or not, is a matter of recent investigations. The {\it Suzaku} X-ray spectrum of 3C 397 showed a strong K-shell emission from stable Fe-peak elements. The high Ni/Fe and Mn/Fe mass ratios in the hot plasma component that dominates the K-shell emission lines may result from exploding white dwarfs with a near-Chandrasekhar mass (\cite{Yamaguchi2015}). The radio location (\cite{Green2014}) of 3C 397 is very close to PSR J1906+0722, which is detected as part of a blind search of unidentified Fermi-LAT sources by the computing system, Einstein@Home\footnote[1]{\url{www.einsteinathome.org}}. This pulsar is associated with the Fermi 3rd catalog source (3FGL) (\cite{Acero2015}), 3FGL J1906.6+0720. For past several years, 3FGL J1906.6+0720 had been known as one of the brightest undefined Fermi-LAT sources but its association was not detected in radio or in X-ray observations conducted before and after the gamma-ray observations. PSR J1906+0722 is a young, energetic, isolated pulsar with a spin frequency of 8.9 Hz, a characteristic age of 49 kyr and spin down power 1.0 $\rm{\times~10^{36}~erg~s^{-1}}$ (\cite{Clark2015}). No pulsar wind nebula (PWN) associated with the PSR J1906+0722, is, however, detected so far (\cite{Xing2014}).
\vspace{-0.5cm}
%%Gamma-ray Analysis
\section{Gamma-ray Analysis}
The large effective area and wide field of view of {\it Large Area Telescope} on board {\it Fermi Gamma Ray Space Telescope} (Fermi-LAT) (\cite{Atwood2009}) allow for the detection and analysis of the gamma-ray emission from pulsars and PWNe. In our analysis, we have selected nearly six years (i.e from 04/08/2008 to 01/10/2014) of Pass 8 source class LAT events in about 100 MeV to 300 GeV energy range within a 20$^{\circ}$ region of interest (ROI) around PSR J1906+0722. We have used \texttt{gtselect} of the {\it Fermi Science Tools} (FST) analysis package\footnote[2]{\url{fermi.gsfc.nasa.gov/ssc/data/analysis/software/}}  to select photons of energies greater than 100 MeV with a 90$^{\circ}$ zenith angle cut to remove contaminations from the Earth's limb.
\begin{wrapfigure}{r}{0.5\textwidth}
\begin{center}
\includegraphics[width=0.48\textwidth]{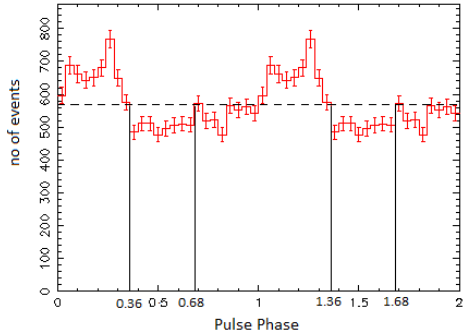}
\end{center}
\caption{{\small The gamma-ray phasogram of PSR J1906+0722.}}
\label{fig1}
\end{wrapfigure}
In order to perform the pulsed analysis of PSR J1906+0722, we have used an updated ephemeris for this pulsar (\cite{Clark2015}). To extract the information about PWN or any other possible associations of the PSR J1906+0722, we have to eliminate the possible contamination due to the strong emissions from the pulsar itself. We have applied the pulsar gating technique\footnote[3]{\url{https://fermi.gsfc.nasa.gov/ssc/data/analysis/scitools/pulsar_gating_tutorial.html}} for such a purpose. Figure \ref{fig1} shows the off-pulse phase interval of PSR J1906+0722 between the phases 0.36 and 0.68. We have then performed the FST binned likelihood analysis\footnote[4]{\url{https://fermi.gsfc.nasa.gov/ssc/data/analysis/scitools/binned_likelihood_tutorial.html}} on this off-pulse phase interval data. For the gamma-ray background model, we have used the diffuse Galactic (\emph{gll$_{-}$iem$_{-}$v6.fits}) and the isotropic (\emph{iso$_{-}$P8R2$_{-}$SOURCE$_{-}\!\!$V6$_{-}\!$v06.txt}) emission components in our analysis. Moreover, we have also incorporated all the 3FGL (\cite{Acero2015}) point sources within a 10$^{\circ}$ $\times$ 10$^{\circ}$ quadratic region centered at the pulsar location. The spectral and normalisation parameters of the sources within 4$^{\circ}$ of the ROI center and the normalizations of the two diffuse background components were left free. After performing this initial analysis, we have added 3C 397 as a new point source with a power-law (PL) type spectrum at the radio location of G41.1-0.3 (\cite{Green2014}) to the background model considered above. We have repeated the analysis with this new background model. The results are summarized in the next section.

\vspace{-0.5cm}
%%Preliminary Results
\section{Preliminary Results}
\vspace{0.1cm}
%3FGL J1906.6+0720 & SNR 3C 397
{{\bf \flushleft {3.1 3FGL J1906.6+0720 \& 3C 397} }}
\vspace{0.1cm}
\par
\begin{figure}
\centering
\includegraphics[width=13cm]{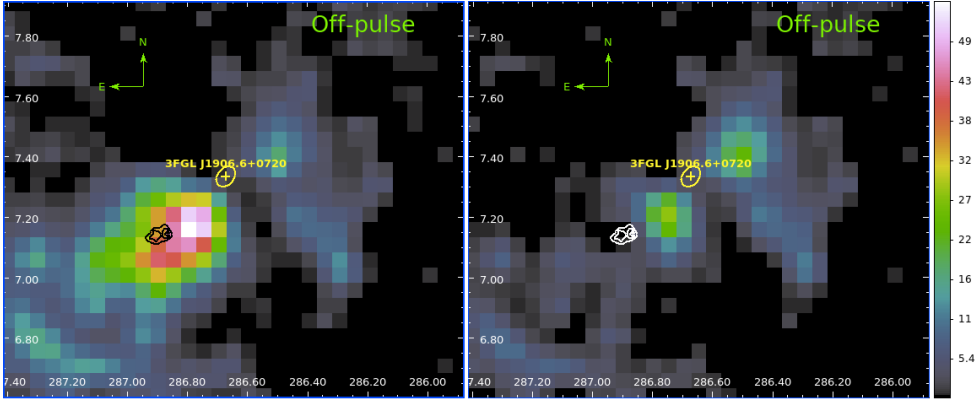} 
\caption{ \footnotesize 10$^{\circ}$ $\times$ 10$^{\circ}$ residual TS map showing the analysis region around the location of 3FGL J1906.6+0720 / PSR J1906+0722. The excess emission near 3C 397 seen in the Left panel has been included in the background model in the Right panel. Contours represent the {\it ROSAT} counts and yellow cross and the circle correspond to the best-fit position and error of 3FGL J1906.6+0720, respectively.}
\label{fig2}
\vspace{-0.3cm}
\end{figure}

 From the binned-likelihood analysis of the off-pulse data, we found that by assuming a Log-parabola type spectrum, the off-pulse emission of 3FGL J1906.6+0720 had a significance of $\sim$31$\sigma$. The spectral indices were found to be $\alpha$ = 2.34 $\pm$ 0.05 and $\beta$ = 0.27 $\pm$ 0.04. The total photon flux and energy flux were obtained to be (4.66 $\pm$ 0.33) $\times$ 10$^{-8}$ photons cm$^{-2}$ sec$^{-1}$ and (3.84 $\pm$ 0.16) $\times$ 10$^{-5}$ MeV cm$^{-2}$ sec$^{-1}$, respectively. Figure \ref{fig2} shows the residual test-statistics (TS) map of the 10$^{\circ}$ $\times$ 10$^{\circ}$ analysis region created by using the off-pulse data, where the off-pulse emission model of 3FGL J1906.6+0720 was included in the background model. While the Left Panel of Figure \ref{fig2} displays the TS map for the case when 3C 397 was not included in the gamma-ray background model, the Right Panel shows the case where the SNR is added as a point source with PL-type spectrum into the background model. The TS difference between the Right and Left Panels leads to the detection of 3C 397 with a significance of $\sim$8$\sigma$. By assuming a PL-type spectrum for 3C 397, we obtained $\Gamma$ = 2.33 $\pm$ 0.09 and the total photon flux and energy flux of 3C 397 were found to be (1.59 $\pm$ 0.30) $\times$ 10$^{-8}$ photons cm$^{-2}$ sec$^{-1}$ and (1.17 $\pm$ 0.16) $\times$ 10$^{-5}$ MeV cm$^{-2}$ sec$^{-1}$, respectively. In Figure \ref{fig2}, the contours of the {\it ROSAT} data are representing the X-ray counts of 8.78, 17.57, 26.36, and 35.15 in black and white colors on the Left and Right Panels, respectively. After modeling out the off-pulse emission of the 3FGL J1906.6+0720 and the excess emission coming from 3C 397, Figure \ref{fig2} shows that there is still some amount of excess in the region  between the pulsar and the SNR. There are various scenarios to explain such excess emission in terms of their possible hadronic or leptonic origins. For example, \cite{Clark2015} suspected that this excess gamma-ray emission could be due to the interaction of 3C 397 with a Molecular Cloud as observed by (\cite{Jiang2010}). 
% PSR J1906+0722 & Possible PWN Association
\vspace{0.1cm}
{{\bf \flushleft {3.2 PSR J1906+0722 \& its Possible PWN Association }}} 
\vspace{0.1cm}
\par
\begin{SCfigure}
\begin{minipage}{9cm}
\flushright
\includegraphics[width=7cm]{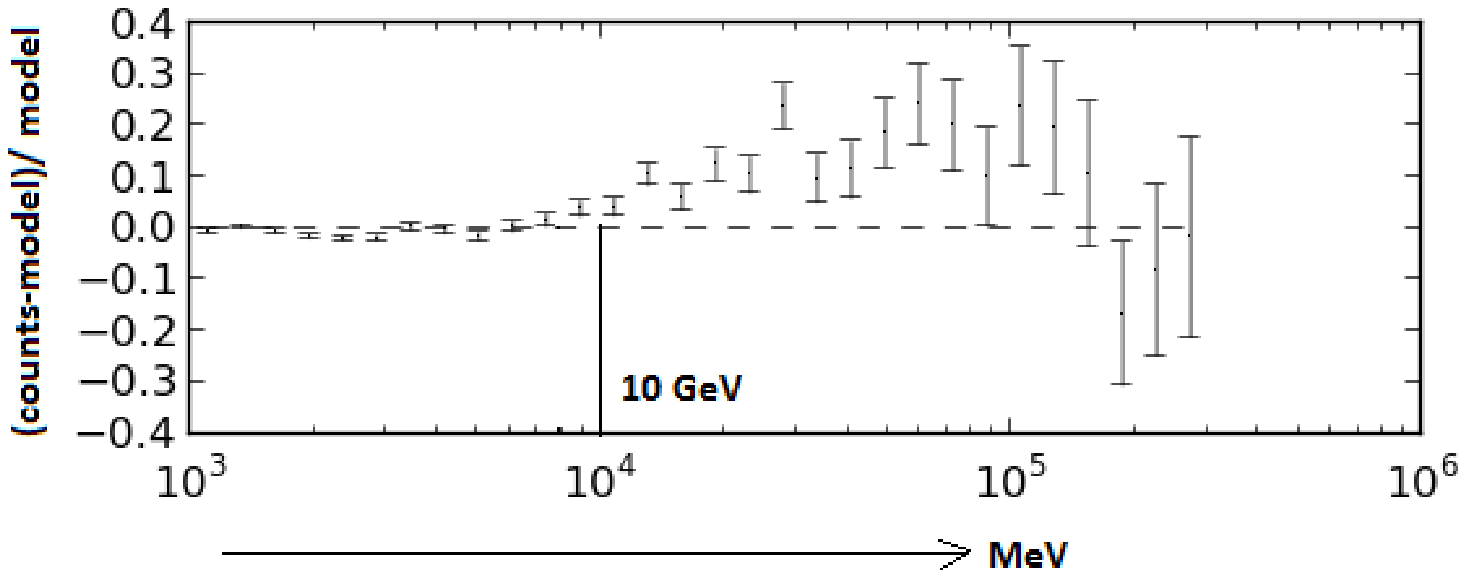}
\includegraphics[width=7cm]{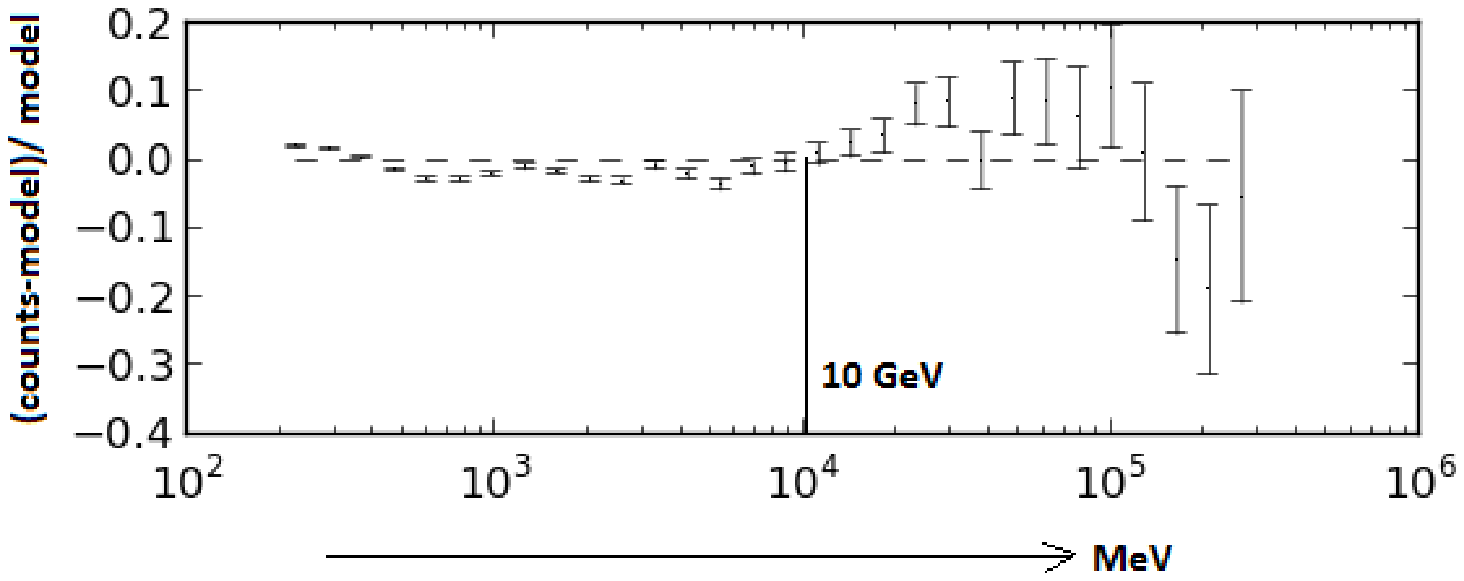}
\end{minipage}
\caption{{\small Residual plots of 10$^{\circ}$ ROI around the location of PSR J1906+0722 shown for two cases: The full-phase data (Upper Panel) and the off-pulse phase data (Lower Panel). The line at 10 GeV is the energy where the excess of gamma-ray emission starts. }}
\label{fig3}
\end{SCfigure}

 Figure \ref{fig3} shows the residual plot of the region around 3FGL J1906.6+0720. In the Upper Panel of this figure, we have displayed the full phase data from PSR J1906+0722. In the Lower panel of Figure \ref{fig3}, we have displayed the off-pulse phase data from PSR J1906+0722, where we applied a phase cut on our dataset. This cut reduces possible contaminations due to strong emissions from the pulsar PSR J1906+0722.
\par It is quite evident from both of the panels of Figure 3 that there is small bump at energies $>$ 10 GeV. It is important
to note that such an excess emission, that appears in the off-pulse phase data from PSR J1906+0722 (displayed in the Lower Panel of Figure 3), is not likely to be related to 3C 397 itself as we have already included 3C 397 in our background model. As PSR J1906+0722 is a young pulsar, it is likely to be associated with a PWN. This excess gamma-ray emission above 10 GeV in Figure \ref{fig3} may, perhaps, be the result of such a possible association of PWN with PSR J1906+0722.
\vspace{-0.8cm}
\section{Conclusion}
\vspace{0.1cm}
We investigated the region of PSR J1906+0722 and 3C 397 in high energy gamma rays by using seventy-four months of Fermi-LAT data. Our preliminary analysis shows that 3FGL J1906.6+0720 has an off-pulse gamma-ray emission of $\sim$31$\sigma$ significance. After eliminating the component due to the pulsar and taking into account the off-pulse emission of PSR J1906+0722 in the background model, we obtained an excess emission of $\sim$8$\sigma$ at the location of 3C 397. After including 3C 397 into the background model, there is still some excess emission left in the region between SNR and the pulsar. Since 3C 397 is supposed to be a Type Ia supernova (\cite{Yamaguchi2015}), it is rather unlikely that it is associated with the pulsar PSR J1906+0722. As seen in Figure \ref{fig2} Right Panel and Figure 3 Lower Panel, we still find some excess gamma-ray emission above 10 GeV in close neighbourhood of the SNR and the pulsar. As PSR J1906+0722 is a young pulsar, there is a possibility that the aforesaid excess emission might arise from an associated PWN. In view of these results, we need to disentangle the SNR from the pulsar and its possibly associated PWN by carefully studying the spectra and the systematic effects. In future, we plan to investigate these points in greater detail.
\vspace{-0.6cm}
%%%ACKNOWLEDGEMENT
\section{Acknowledgement} 
\footnotesize {The authors are grateful to Dr. C. J. Clark, Albert-Einstein-Institut, MPIG, Germany for making the Ephemeris of PSR J1906+0722 accessible to us. PB is particularly grateful to the International Astronomical Union (IAU) for the allocation of a grant thus enabling her to attend this IAU S331 symposium. We are grateful to all the organizers of the IAUS331 symposium for their warm hospitality. PB is supported by DST INSPIRE FELLOWSHIP. TE thanks the Science Academy, Turkey for the support under the Science Academy's Young Scientists Award Program (BAGEP-2015), Turkey. We would like to thank the referee for carefully reading the manuscript and for giving very constructive comments.}
\vspace{-0.7cm}
%%%BIBLIOGRAPHY

%%%END OF DOCUMENT
\end{document}